# On Four-group ML Decodable Distributed Space Time Codes for Cooperative Communication

G. Susinder Rajan, Anshoo Tandon and B. Sundar Rajan
Department of Electrical Communication Engineering
Indian Institute of Science, Bangalore, India
Email: {susinder, anshoo, bsrajan}@ece.iisc.ernet.in

*Abstract*— A construction of a new family of distributed space time codes (DSTCs) having full diversity and low Maximum Likelihood (ML) decoding complexity is provided for the two phase based cooperative diversity protocols of Jing-Hassibi and the recently proposed Generalized Non-orthogonal Amplify and Forward (GNAF) protocol of Rajan et al. The salient feature of the proposed DSTCs is that they satisfy the extra constraints imposed by the protocols and are also four-group ML decodable which leads to significant reduction in ML decoding complexity compared to all existing DSTC constructions. Moreover these codes have uniform distribution of power among the relays as well as in time. Also, simulations results indicate that these codes perform better in comparison with the only known DSTC with the same rate and decoding complexity, namely the Coordinate Interleaved Orthogonal Design (CIOD). Furthermore, they perform very close to DSTCs from field extensions which have same rate but higher decoding complexity.

I. INTRODUCTION

Cooperative diversity is a technique by which multiple terminals (users or relays) cooperate to form a virtual antenna array thereby leveraging the spatial diversity benefits even if a local antenna array is not available. A cooperative diversity protocol dictates how the users would actually cooperate among themselves to achieve the required diversity order. Several cooperative diversity protocols have been proposed in the literature [1]-[6]. In this paper, we focus on the two phase based protocols of Jing-Hassibi [1] and the GNAF protocol of Rajan et al [2] for three reasons- (i) the operations at the relay nodes are considerably simplified, (ii) we can avoid imposing bottlenecks on the rate by not requiring the relay nodes to decode and (iii) the framework of distributed space-time codes allows for more flexibility and higher spectral efficiency [1], [2], [4]. Transmission in this protocol comprises of two phases- broadcast phase and cooperation phase. In the broadcast phase, the source broadcasts its information to the relays and the destination. In the cooperation phase, each relay transmits a linearly processed version of the received vector. For this purpose, each relay is equipped with a unitary matrix which we call 'relay matrix'. It was shown in [1], [2] that to the destination it would appear as if a space-time code was transmitted from colocated multiple antennas. Further the design criteria to achieve full diversity also remains to be the well known rank criteria for colocated MIMO (Multiple Input Mutiple Output Systems). However it is important to note that DSTCs need to satisfy many additional constraints on the code structure (for example, unitary relay matrices) and are different from the traditional Space-time codes for colocated MIMO systems, which are designed without respecting such constraints.

The ML decoding of a STBC in $K$ complex variables $x_1, x_2, \cdots, x_K$ is, in general, joint decoding of all the $K$ variables. However, for the Alamouti code $K = 2$ and the variables $x_1$ and $x_2$ can be decoded independently for ML decoding. In general, if $K = g\lambda$ and the variables can be partitioned into $g$ subsets each containing $\lambda$ number of variables and the ML decoding can be done for the variables of a subset independently of the variables of other subsets the code is said to be $g$-group ML decodable or $\lambda$-symbol decodable [9], [10]. The Alamouti code is single-symbol decodable or two-group ML decodable. Following the work of [1], several distributed space time codes [6], [7], [8] were proposed. However, most of these code constructions did not consider the important aspect of ML decoding complexity at the destination. This problem gains significant importance especially if *the number of relays in the network is large*. An initiative in this direction was first taken in [11] wherein two-group ML decodable DSTCs were proposed using a construction procedure called 'doubling construction'. Later in [3], a class of rate one, full diversity, four-group ML decodable DSTCs called Precoded CIODs(Co-ordinate Interleaved Orthogonal Designs) were proposed for arbitrary number of relays. However these DSTCs had a large number of zero entries in the design which led to a large Peak to Average Power Ratio (PAPR). In [10], a class of four-group ML decodable STCs were proposed for the colocated MIMO systems. However these STCs fail to satisfy the additional constraints on the code structure imposed by the cooperative diversity protocol. Hence it is important to address the problem of constructing a new class of four-group ML decodable DSTCs which have low PAPR and uniform distribution of power among the relays.

The main contribution of this paper is in providing a new construction of DSTCs with the following salient features.

- Source can transmit $\frac{1}{2}$ complex symbols per channel use
- Full diversity
- Four group ML decodable
- Unitary relay matrices. (This eliminates the need for a whitening filter at the destination thus further reducing decoding complexity.)
- Unitary weight matrices together with unitary relay matrices makes the power distribution uniform among the

relays and in time and hence results in low PAPR.

The paper is organized as follows: We briefly describe the notion of four-group ML decodable codes and also state the DSTC design constraints which are due to the protocol in Section II. In Section III, we describe the code construction procedure and show that they possess the salient features stated above. Few illustrative examples of DSTC construction are also provided for 4 and 16 relays. Simulation results are presented in Section IV. Conclusions and discussions on further work comprise Section V.

**Notation:** For a complex matrix $A$, $A^*$, $A^T$ and $A^H$ denote the conjugate, transpose and conjugate transpose respectively. $A_I$ denotes the real matrix obtained by taking the real parts of all the entries of the matrix $A$ and $A_Q$ denotes the real matrix obtained by taking the imaginary parts of all the entries of the matrix $A$. For a square matrix $B$, $|B|$ and $\text{Tr}(B)$ denote the determinant and trace of the matrix $B$ respectively.

## II. FOUR GROUP ML DECODABLE DSTC DESIGN PROBLEM

In this section, we briefly describe the two phase based cooperative diversity protocol of [1], [2] and introduce the problem statement. Consider a wireless relay network consisting of a source node, a destination node and $R$ other relay nodes which aid the source in communicating information to the destination. The channel path gains from the source to the $i^{th}$ relay, denoted by $f_i$ and those from the $j^{th}$ relay to the destination denoted by $g_j$ are all assumed to be i.i.d $\mathcal{CN}(0,1)$. The channel path gain, $g_0$ from the source to the destination is also assumed to be $\mathcal{CN}(0,1)$. All the nodes are equipped with a single antenna and are subject to the half-duplex constraint, i.e., a node cannot transmit and receive simultaneously. Further, we assume that the nodes are synchronized at the symbol level. Each transmission from source to destination comprises of two phases- broadcast phase and cooperation phase. In the broadcast phase, the source transmits a $T$ length vector $s$ taken from a codebook consisting of information vectors $\mathscr{C} = \{s_1, \ldots, s_L\}$ satisfying $\text{E}\{s^H s\} = 1$ to all the relays and the destination. In the cooperation phase, all the relay nodes are scheduled to transmit together (assuming symbol level synchronization) a DSTC. For this purpose, each relay is equipped with a unitary matrix $A_i$ which we call 'relay matrix'. To be precise, the $i^{th}$ relay transmits a scaled version (to satisfy power constraint) of $A_i r_i$ or $A_i r_i^*$, where $r_i$ denotes the received vector at the relay. The signal model is as shown in (1) at the top of the next page, where

- $t_i$ denotes the vector transmitted by the $i^{th}$ relay and $v_i$ denotes the additive noise at the relay whose entries are assumed to be i.i.d $\mathcal{CN}(0,1)$.
- $y_{D,1}$ and $y_{D,2}$ denote the received vector at the destination during the broadcast phase and cooperation phase respectively. $w_1$ and $w_2$ represent the additive noise at the destination whose entries are i.i.d $\mathcal{CN}(0,1)$. The quantities $\pi_1$ and $\pi_2$ are the power allocation factors satisfying $\pi_1 + \pi_2 TR = 2T$ so that $P$ represents the total average power spent by the source and the relays together.

The received vector at the destination can be written in matrix form as follows

$$y = \begin{bmatrix} y_{D,1} \\ y_{D,2} \end{bmatrix} = \sqrt{\frac{\pi_2 \pi_1 P^2}{\pi_1 P + 1}} SH + W \quad (2)$$

where $S, H$ and $W$ are as shown in (3) at the top of the next page. The DSTC in this case is the collection of all the $(2T) \times (R+1)$ matrices $S$. Observing the structure of $S$ in (3), we see that it is sufficient to design the submatrix of $S$ given by $S_{ER} = \begin{bmatrix} A_1 s & \ldots & A_N s & A_{N+1} s^* & \ldots & A_R s^* \end{bmatrix}$. We are interested in the case when the submatrix $S_{ER}$ is obtained from a linear dispersion STBC say $S(X)$. Let $S(X) = \sum_{i=1}^{K} x_i A_i$ where, $x_1, x_2, \ldots, x_K$ are the $K$ real variables of the linear STBC $S(X)$ and the matrices $A_i \in \mathbb{C}^{T \times N_t}$, called the 'weight matrices', define the code.

The vector $X = [x_1, x_2, \ldots, x_K]^T \in \mathscr{A} \subset \mathbb{R}^K$ is called the information symbol vector.

Suppose we partition the information symbol vector as $X^T = [X_1^T, X_2^T, \ldots, X_g^T]$, where $X_k^T = [x_{j_k+1}, x_{j_k+2}, \ldots, x_{j_k+n_k}]$, $j_1 = 0$ and $j_k = \sum_{i=1}^{k-1} n_i$ for $k = 1, 2, \ldots, g$ and their corresponding set of weight matrices also into $g$ groups $L_k, k = 1, \ldots, g$, the $k^{th}$ group containing $n_k$ matrices, then $S(X)$ can be written as

$$S(X) = \sum_{k=1}^{g} S_k(X_k), \text{where} \quad S_k(X_k) = \sum_{i=1}^{n_k} x_{j_k+i} A_{j_k+i}.$$

Now if the information symbols in each group take values independent of information symbols in the other groups and if the weight matrices satisfy

$$A_i^H A_j + A_j^H A_i = \mathbf{0}, \quad \forall i \in L_p, \forall j \in L_q, p \neq q. \quad (4)$$

then, it can be easily shown [9], [10] that $S(X)$ is $g$-group ML decodable. In other words, the ML decoding can be performed by minimizing the metric

$$\| y - \sqrt{\frac{\pi_2 \pi_1 P^2}{\pi_1 P + 1}} S_k(X_k) H \|^2 \quad (5)$$

for each $1 \leq k \leq g$ individually instead of minimizing the computationally more intensive metric

$$\| y - \sqrt{\frac{\pi_2 \pi_1 P^2}{\pi_1 P + 1}} S(X) H \|^2 \quad (6)$$

Hence the ML decoding complexity is reduced to a large extent depending on the value of $g$. In particular, the decoding complexity is reduced from $2^\lambda$ to $g 2^{\frac{\lambda}{g}}$ where, $\lambda$ depends on the rate of transmission as measured in bits/s/Hz. In this paper, we consider only the case of $g = 4$. Combining all the given requirements, the 4-group ML decodable DSTC design problem is thus to find space-time block codes satisfying the following three constraints.

1) *Any column should contain only the variables of the design or only their conjugates.*

$$\begin{aligned}
y_{D,1} &= \sqrt{\pi_1 P} g_0 s + w_1 \\
r_i &= \sqrt{\pi_1 P} f_i s + v_i, \forall \quad i = 1, \ldots, R \\
t_i &= \sqrt{\frac{\pi_2 P}{\pi_1 P + 1}} A_i r_i, \quad i = 1, \ldots, N \\
t_i &= \sqrt{\frac{\pi_2 P}{\pi_1 P + 1}} A_i r_i^*, \quad i = (N+1), \ldots, R \\
y_{D,2} &= \sum_{i=1}^{R} g_i t_i + w_2
\end{aligned} \quad (1)$$

$$\begin{aligned}
S &= \begin{bmatrix} \sqrt{\frac{\pi_1 P + 1}{\pi_2 P}} I_{T_1} s & 0 & \ldots & 0 & 0 & \ldots & 0 \\ 0 & A_1 s & \ldots & A_N s & A_{N+1} s^* & \ldots & A_R s^* \end{bmatrix} \\
H^T &= \begin{bmatrix} g_0 & g_1 f_1 & \ldots & g_N f_N & g_{N+1} f_{N+1}^* & \ldots & g_R f_R^* \end{bmatrix} \\
W &= \begin{bmatrix} w_1 \\ \sqrt{\frac{\pi_2 P}{\pi_1 P + 1}} \left( \sum_{i=1}^N g_i A_i v_i + \sum_{i=N+1}^R g_i A_i v_i^* \right) + w_2 \end{bmatrix}
\end{aligned} \quad (3)$$

2) *All the relay matrices should be unitary.*
3) *The weight matrices of the code should satisfy the conditions for $4$-group decodability.*

### III. CODE CONSTRUCTION PROCEDURE

Having described the problem statement, in this section we explicitly construct a new class of rate one, full diversity four-group ML decodable DSTCs.

Consider the following design of size $R \times R$,

$$S = \begin{bmatrix} A & -B^H \\ B & A^H \end{bmatrix} \quad (7)$$

where, $A$ and $B$ are identical $\frac{R}{2} \times \frac{R}{2}$ designs in different variables. If the codewords in $A$ commute with the codewords in $B$, then we have

$$S^H S = \begin{bmatrix} A^H A + B^H B & 0 \\ 0 & BB^H + AA^H \end{bmatrix}. \quad (8)$$

This ensures that the code $S$ is two-group ML decodable, one group involving variables in $A$ and the other group involving the variables in $B$. This fact was exploited in [11] to construct two-group ML decodable DSTCs. This construction procedure was named as 'Doubling construction' in [11]. However, note that in addition if $A$ and $B$ are individually two-group ML decodable, then it can be easily shown that the code $S$ will be four-group ML decodable. In this paper, we essentially exploit this fact. Hence we require two-group ML decodable codes whose codewords commute among themselves. Towards that end, consider the following $2 \times 2$ designs.

$$C_1 = \begin{bmatrix} s_1 & s_2 \\ -s_2 & s_1 \end{bmatrix} \text{ and } C_2 = \begin{bmatrix} s_1 & s_2 \\ s_2 & s_1 \end{bmatrix}$$

We can show that both $C_1$ and $C_2$ are two-group ML decodable codes. For the design $C_2$, the two groups are $\{s_{1I}, s_{2Q}\}$ and $\{s_{1Q}, s_{2I}\}$ respectively. More importantly, we have the following identity true for any complex number $\gamma$.

$$\begin{bmatrix} s_1 & s_2 \\ \gamma s_2 & s_1 \end{bmatrix} \begin{bmatrix} s_1' & s_2' \\ \gamma s_2' & s_1' \end{bmatrix} = \begin{bmatrix} s_1' & s_2' \\ \gamma s_2' & s_1' \end{bmatrix} \begin{bmatrix} s_1 & s_2 \\ \gamma s_2 & s_1 \end{bmatrix}$$

In other words, codewords arising from design $C_1$ or $C_2$ commute. Thus we can construct a $4 \times 4$ design by substituting either $C_1$ or $C_2$ for $A$ and $B$ in (7) as shown below.

$$S = \begin{bmatrix} s_1 & s_2 & -s_3^* & -s_4^* \\ s_2 & s_1 & -s_4^* & -s_3^* \\ s_3 & s_4 & s_1^* & s_2^* \\ s_4 & s_3 & s_2^* & s_1^* \end{bmatrix} \quad (9)$$

It can be easily verified that $S$ is indeed 4-group ML decodable. The four-groups are $\{s_{1I}, s_{2I}\}$, $\{s_{1Q}, s_{2Q}\}$, $\{s_{3I}, s_{4I}\}$ and $\{s_{3Q}, s_{4Q}\}$ respectively. To extend this approach for more number of relays, we need to find designs analogous to $C_1$ and $C_2$ for higher dimensions. Towards that end, consider the $\frac{R}{2} \times \frac{R}{2}$ design

$$D = \begin{bmatrix} W & X \\ X & W \end{bmatrix}$$

where, $W$ and $X$ are identical $\frac{R}{4} \times \frac{R}{4}$ designs in different variables. We call this construction as the 'ABBA' construction. It is easy to check that if the design $W$ or $X$ is $g$-group ML decodable, then the design $D$ obtained using ABBA construction is also $g$-group ML decodable. Further, we have

$$\begin{bmatrix} W & X \\ X & W \end{bmatrix} \begin{bmatrix} W' & X' \\ X' & W' \end{bmatrix} = \begin{bmatrix} WW' + XX' & WX' + XW' \\ XW' + WX' & XX' + WW' \end{bmatrix}$$

$$\begin{bmatrix} W' & X' \\ X' & W' \end{bmatrix} \begin{bmatrix} W & X \\ X & W \end{bmatrix} = \begin{bmatrix} W'W + X'X & W'X + X'W \\ X'W + W'X & X'X + W'W \end{bmatrix}.$$

Thus we observe that codewords obtained from design $D$ commute if $W'W = WW'$, $X'X = XX'$, $W'X = XW'$ and $X'W = WX'$. Since $W$ and $X$ are taken to be identical designs, this simply means that if codewords in $W$ commute, then codewords of $D$ also commute. Using the above facts, we now give our construction procedure for any $R = 2^a$ relays for $a > 2$.

*Construction 1:* The steps in the construction procedure are enlisted as below.

Step 1: Starting with either $C_1$ or $C_2$, keep applying ABBA construction iteratively on it till a $\frac{R}{2} \times \frac{R}{2}$ design $C$ is obtained. The design $C$ will be two-group ML decodable and it will also

$$S = \begin{bmatrix}
s_1 & s_2 & s_3 & s_4 & s_5 & s_6 & s_7 & s_8 & -s_9^* & -s_{10}^* & -s_{11}^* & -s_{12}^* & -s_{13}^* & -s_{14}^* & -s_{15}^* & -s_{16}^* \\
s_2 & s_1 & s_4 & s_3 & s_6 & s_5 & s_8 & s_7 & -s_{10}^* & -s_9^* & -s_{12}^* & -s_{11}^* & -s_{14}^* & -s_{13}^* & -s_{16}^* & -s_{15}^* \\
s_3 & s_4 & s_1 & s_2 & s_7 & s_8 & s_5 & s_6 & -s_{11}^* & -s_{12}^* & -s_9^* & -s_{10}^* & -s_{15}^* & -s_{16}^* & -s_{13}^* & -s_{14}^* \\
s_4 & s_3 & s_2 & s_1 & s_8 & s_7 & s_6 & s_5 & -s_{12}^* & -s_{11}^* & -s_{10}^* & -s_9^* & -s_{16}^* & -s_{15}^* & -s_{14}^* & -s_{13}^* \\
s_5 & s_6 & s_7 & s_8 & s_1 & s_2 & s_3 & s_4 & -s_{13}^* & -s_{14}^* & -s_{15}^* & -s_{16}^* & -s_9^* & -s_{10}^* & -s_{11}^* & -s_{12}^* \\
s_6 & s_5 & s_8 & s_7 & s_2 & s_1 & s_4 & s_3 & -s_{14}^* & -s_{13}^* & -s_{16}^* & -s_{15}^* & -s_{10}^* & -s_9^* & -s_{12}^* & -s_{11}^* \\
s_7 & s_8 & s_5 & s_6 & s_3 & s_4 & s_1 & s_2 & -s_{15}^* & -s_{16}^* & -s_{13}^* & -s_{14}^* & -s_{11}^* & -s_{12}^* & -s_9^* & -s_{10}^* \\
s_8 & s_7 & s_6 & s_5 & s_4 & s_3 & s_2 & s_1 & -s_{16}^* & -s_{15}^* & -s_{14}^* & -s_{13}^* & -s_{12}^* & -s_{11}^* & -s_{10}^* & -s_9^* \\
s_9 & s_{10} & s_{11} & s_{12} & s_{13} & s_{14} & s_{15} & s_{16} & s_1^* & s_2^* & s_3^* & s_4^* & s_5^* & s_6^* & s_7^* & s_8^* \\
s_{10} & s_9 & s_{12} & s_{11} & s_{14} & s_{13} & s_{16} & s_{15} & s_2^* & s_1^* & s_4^* & s_3^* & s_6^* & s_5^* & s_8^* & s_7^* \\
s_{11} & s_{12} & s_9 & s_{10} & s_{15} & s_{16} & s_{13} & s_{14} & s_3^* & s_4^* & s_1^* & s_2^* & s_7^* & s_8^* & s_5^* & s_6^* \\
s_{12} & s_{11} & s_{10} & s_9 & s_{16} & s_{15} & s_{14} & s_{13} & s_4^* & s_3^* & s_2^* & s_1^* & s_8^* & s_7^* & s_6^* & s_5^* \\
s_{13} & s_{14} & s_{15} & s_{16} & s_9 & s_{10} & s_{11} & s_{12} & s_5^* & s_6^* & s_7^* & s_8^* & s_1^* & s_2^* & s_3^* & s_4^* \\
s_{14} & s_{13} & s_{16} & s_{15} & s_{10} & s_9 & s_{12} & s_{11} & s_6^* & s_5^* & s_8^* & s_7^* & s_2^* & s_1^* & s_4^* & s_3^* \\
s_{15} & s_{16} & s_{13} & s_{14} & s_{11} & s_{12} & s_9 & s_{10} & s_7^* & s_8^* & s_5^* & s_6^* & s_3^* & s_4^* & s_1^* & s_2^* \\
s_{16} & s_{15} & s_{14} & s_{13} & s_{12} & s_{11} & s_{10} & s_9 & s_8^* & s_7^* & s_6^* & s_5^* & s_4^* & s_3^* & s_2^* & s_1^*
\end{bmatrix} \quad (10)$$

have commuting codewords.

Step 2: Apply doubling construction (7) to $C$ by substituting $C$ for $A$ and $B$(albeit with different variables). Thus we obtain a $R \times R$ design which will be four-group ML decodable.

Note that the codes designed using Construction 1 have the property that *any column has only the variables or only their conjugates*. Moreover it can be shown that these codes have *unitary relay matrices as well as unitary weight matrices*. This leads to uniform distribution of power among the relays as well as in time and hence low PAPR.

*Example 1:* Using construction 1, we get the $16 \times 16$ design shown in (10) at the top of this page. The four groups of variables that can be decoded separately are $\{s_{1I}, s_{2I}, \ldots, s_{8I}\}$, $\{s_{1Q}, s_{2Q}, \ldots, s_{8Q}\}$, $\{s_{9I}, s_{10I}, \ldots, s_{16I}\}$ and $\{s_{9Q}, s_{10Q}, \ldots, s_{16Q}\}$ respectively. Observe that any column has only the variables or only its conjugates. Further note that all the relay matrices are unitary. It can also be checked that all the weight matrices are unitary.

### A. Diversity of constructed codes

The constructed codes can be made to achieve full diversity by letting the variables in each of the four groups to take values from an appropriately rotated $\mathbb{Z}^{\frac{R}{2}}$ lattice constellation. We would like to emphasize here that the variables in a group should be allowed to take values independent of the variables in the other groups. Algebraic number theory provides effective means to construct rotated $\mathbb{Z}^n$ lattices with full diversity and large minimum product distance [12]. Such lattices can be used to obtain full diversity for our codes. We will now illustrate how this can be done for the $4 \times 4$ design shown in (9).

Note that the matrix $S$ is invertible if either $A$ or $B$ is invertible. This is because,

$$\begin{aligned} |S| &= |S^H S|^{\frac{1}{2}} = |(A^H A + B^H B)(AA^H + BB^H)|^{\frac{1}{2}} \\ &\geq \max\{|A|^2, |B|^2\} \end{aligned} \quad (11)$$

where, the second inequality was obtained on application of Corollary 4.3.3 in [13]. Therefore it is sufficient to show how $C_2$ can be made fully diverse. We have

$$|\Delta C_2| = (\Delta s_1 + \Delta s_2)(\Delta s_1 - \Delta s_2). \quad (12)$$

Therefore the condition for full-diversity is $\Delta s_1 \neq \pm \Delta s_2$. Thus, if $\Delta s_{1I} \neq \pm \Delta s_{2I}$ and $\Delta s_{1Q} \neq \pm \Delta s_{2Q}$ then full diversity is guaranteed. Let us define $p = \frac{1}{\sqrt{2}}(s_{1I} + s_{2I})$ and $q = \frac{1}{\sqrt{2}}(s_{1I} - s_{2I})$. Then the required condition is simply $\Delta p \Delta q \neq 0$, which is nothing but non zero product distance. So let $p, q$ take values from a rotated $\mathbb{Z}^2$ lattice constellation designed to maximize product distance [12]. Then let $s_{1I} = \frac{p+q}{\sqrt{2}}$ and $s_{2I} = \frac{p-q}{\sqrt{2}}$. Similarly it is done for $s_{1Q}, s_{2Q}$ also. Thus we have

$$\begin{bmatrix} s_{1I} \\ s_{2I} \end{bmatrix} = \begin{bmatrix} \frac{1}{\sqrt{2}} & \frac{1}{\sqrt{2}} \\ \frac{1}{\sqrt{2}} & -\frac{1}{\sqrt{2}} \end{bmatrix} G \mathbb{Z}^2 \quad (13)$$

$$\begin{bmatrix} s_{1Q} \\ s_{2Q} \end{bmatrix} = \begin{bmatrix} \frac{1}{\sqrt{2}} & \frac{1}{\sqrt{2}} \\ \frac{1}{\sqrt{2}} & -\frac{1}{\sqrt{2}} \end{bmatrix} G \mathbb{Z}^2 \quad (14)$$

where, $G = \begin{bmatrix} -0.5257311121 & -0.8506508083 \\ -0.8506508083 & 0.5257311121 \end{bmatrix}$.

In this manner, we can show that it is possible to achieve full diversity for all the codes in this paper.

### IV. SIMULATION RESULTS

In this section, we compare the performance of the newly proposed DSTC for a 4 relay network (9) with that of DSTCs from CIOD [2] and field extensions [6], [7], [8]. The power allocation factors used for the simulation are $\pi_1 = T$, and $\pi_2 = \frac{1}{R}$. The signal sets chosen for the different DSTCs are as follows:

1) CIOD - QPSK rotated by $31.7175°$
2) Field extension - QPSK
3) Newly proposed code - QPSK appropriately rotated as explained in subsection A of Section III.

Fig.1 shows the codeword error rate comparison of the newly proposed DSTC with that of DSTCs from CIOD and field extensions. Observe from Fig.1 that the newly proposed code performs slightly better compared to the 4-group ML decodable CIOD [2]. Moreover its performance is only slightly inferior compared to the 1-group ML decodable DSTC from field extensions [6], [7], [8].

### V. DISCUSSION

A class of full diversity 4-group ML decodable DSTCs with unitary relay matrices and unitary weight matrices were explicity constructed in this paper for number of relays of the form $R = 2^a, a \in \mathbb{N}$. The effect of power allocation on the error performance was not considered in this work and is an interesting problem for further work. Further study on the codes of this paper has revealed an interesting algebraic structure which gives a unifying theory of several known codes in the literature including the ones in this paper. An algebraic approach to designing these codes is currently under progress.

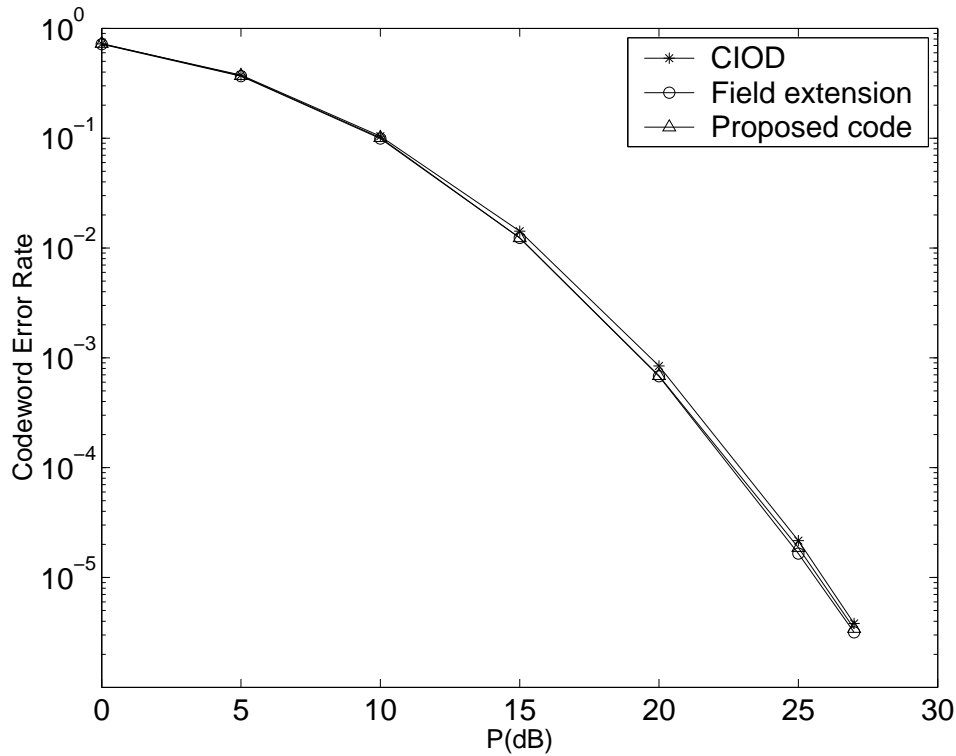

Fig. 1. Performance comparison of the proposed code with DSTCs from CIOD and field extension

ACKNOWLEDGMENT

This work was partly supported by the DRDO-IISc Program on Advanced Research in Mathematical Engineering, partly by the Council of Scientific & Industrial Research (CSIR), India, through Research Grant (22(0365)/04/EMR-II) to B.S. Rajan. The authors are grateful to the anonymous reviewers for their valuable comments which helped improve the clarity of the paper.

REFERENCES

[1] Yindi Jing and Babak Hassibi, "Distributed Space-Time Coding in Wireless Relay Networks", *IEEE Transactions on Wireless Communications*, Vol.5, No.12, pp.3524 - 3536, December 2006.
[2] G.Susinder Rajan and B.Sundar Rajan, "A Non-orthogonal distributed space-time protocol, Part-I: Signal model and design criteria," Proceedings of *IEEE International Workshop in Information Theory*, Chengdu, China, Oct.22-26, 2006, pp.385-389.
[3] ———, "A Non-orthogonal distributed space-time protocol, Part-II: Code construction and DM-G Tradeoff," Proceedings of *IEEE International Workshop in Information Theory*, Chengdu, China, Oct.22-26, 2006, pp.488-492.
[4] J. N. Laneman and G. W. Wornell, "Distributed Space-Time Coded Protocols for Exploiting Cooperative Diversity in Wireless Networks," *IEEE Trans. Inform. Theory*, Vol. 49, No. 10, pp. 2415-2525, Oct. 2003. "Cooperative Diversity in Wireless Networks: Efficient Protocols and Outage Behavior," *IEEE Trans. Inform. Theory*, Vol. 50, No. 12, pp. 3062-3080, Dec. 2004.
[5] K. Azarian, H. El Gamal, and P. Schniter, "On the achievable diversity multiplexing tradeoff in half-duplex cooperative channels," *IEEE Trans. Inform. Theory*, Vol.51, No.12, Dec. 2005, pp.4152-4172.
[6] Petros Elia, P. Vijay Kumar, "Diversity-Multiplexing Optimality and Explicit Coding for Wireless Networks With Reduced Channel Knowledge," Presented at the 43rd Allerton Conference on Communications, Control and Computing, Sept. 2005.
[7] Frederique Oggier, Babak Hassibi, "An Algebraic Family of Distributed Space-Time Codes for Wireless Relay Networks," Proc. *IEEE International Symposium on Information Theory*, Seattle, July 9-14, 2006, pp.538-541.
[8] Petros Elia, Frederique Oggier and P. Vijay Kumar, Asymptotically Optimal Cooperative Wireless Networks without Constellation Expansion, Proc. *IEEE International Symposium on Information Theory*, Seattle, July 9-14, 2006, pp.2729-2733.
[9] Md. Zafar Ali Khan and B. Sundar Rajan, "Single-Symbol Maximum-Likelihood Decodable Linear STBCs," *IEEE Trans. on Inform. Theory*, Vol.52, No.5, pp.2062-2091, May 2006.
[10] Chau Yuen, Yong Liang Guan, Tjeng Thiang Tjhung, "A class of four-group quasi-orthogonal space-time block code achieving full rate and full diversity for any number of antennas," in Proceedings *IEEE Personal, Indoor and Mobile Radio Communications Symposium*(PIMRC), Vol.1, pp.92 - 96, Berlin, Germany, Sept.11-14, 2005.
[11] Kiran T. and B. Sundar Rajan, "Distributed Space-Time Codes with Reduced Decoding Complexity," Proceedings of *IEEE International Symposium on Inform. Theory*, Seattle, July 9-14, 2006, pp. 542-546.
[12] Full Diversity Rotations, http://www1.tlc.polito.it/∼viterbo/rotations/rotations.html.
[13] R.A.Horn and C.R.Johnson, Matrix Analysis. Cambridge, U.K.: Cambridge Univ. Press, 1985.